\input{aipcheck}

\documentclass[sort
    ,final            
  ]
  {aipproc}

\layoutstyle{6x9}

\begin{document}

\title{Unraveling the dynamics and kinematics of GRB hosts with high resolution spectroscopy}


\author{Christina C. Th\"one}{
  address={Dark Cosmology Centre, Juliane Maries Vej 30, 2100 Copenhagen, Denmark}}
  \author{Johan P. U. Fynbo}{
  address={Dark Cosmology Centre, Juliane Maries Vej 30, 2100 Copenhagen, Denmark}}
 \author{Lise Christensen}{
 address={European Southern Observatory, Casilla 19001, Santiago 19, Chile}}
 \author{Klaas Wiersema}{
  address={Department of Physics and Astronomy, Univ. of
Leicester, University Road, Leicester, UK}}
  \author{Joshua S. Bloom}{
  address={University of California Berkeley, 601 Campbell Hall, Berkeley, CA 94720, USA}
}

\begin{abstract}
In the last years the research on GRB host galaxies has proceeded to more detailed studies  both using high resolution afterglow spectroscopy and spatially resolved spectra of nearby hosts. High resolution spectra give a detailed picture of the kinematic properties of matter along the line-of-sight in the host. Two afterglow spectra show clear indications for outflows from their host, namely GRB\,030329 and GRB\,060206, derived from the position of absorption and host emission lines, nonvariability of Mg I and the radiation field calculated from fine-structure line detections. In nearby GRB hosts it is possible to resolve the actual GRB site. GRB\,060505, a SN-less GRB, originated in a relatively metal poor, star forming region with similar properties as other long-duration hosts. A similar conclusion is reached for the site of GRB\,980425/SN 1998bw.
 \end{abstract}

\maketitle


Host galaxies of GRBs have been studied since the discovery of the first afterglow in 1997 providing a subarcsecond localization of the burst and the subsequent identification with an underlying host. About 100 host galaxies of long and 11 for short GRBs have been identified to date. Long duration GRB hosts are a rather uniform population being young ($<$100 Myr), actively star forming ($>$5--10 M$_\odot$/yr), subluminous (0.1 L*) galaxies \citep{Christensen04, Fruchter06} with some exceptions of  late-type spirals where the GRB occurred in star-forming regions in a spiral arm.\\
A detailed look on the stellar population the GRB was born into is usually hampered by the lack of spatial resolution except for some nearby hosts. Nevertheless, for high redshift bursts, we can study the properties of the ISM in the line of sight, where high-resolution spectroscopy has provided insight into the kinematics of the ISM, and the fine-structure lines excited by UV pumping allows us to derive distances to the burst \citep{Prochaska07}. This proceeding presents four bursts studied with high resolution and/or spatially resolved spectroscopy.


\section{High resolution spectra - the distant universe}
Most absorption systems detected show small velocity spreads of a few hundred km/s only resolved in high resolution spectra. Comparable structures have been detected in QSO absorbers \citep[e.g.][]{Ledoux06, Murphy07, Prochaska08} and Lyman Break Galaxies (LBGs) \citep{Pettini02}. They might either be caused by material in the disc/halo of the galaxy \citep{Ellison} or by cold clumps in a starburst wind where cumulative supernova explosions drive metal enriched matter into the IGM. Similar structures were detected in NaD and OVI lines in nearby starbursts.

\begin{figure}
\includegraphics[height=8.3cm]{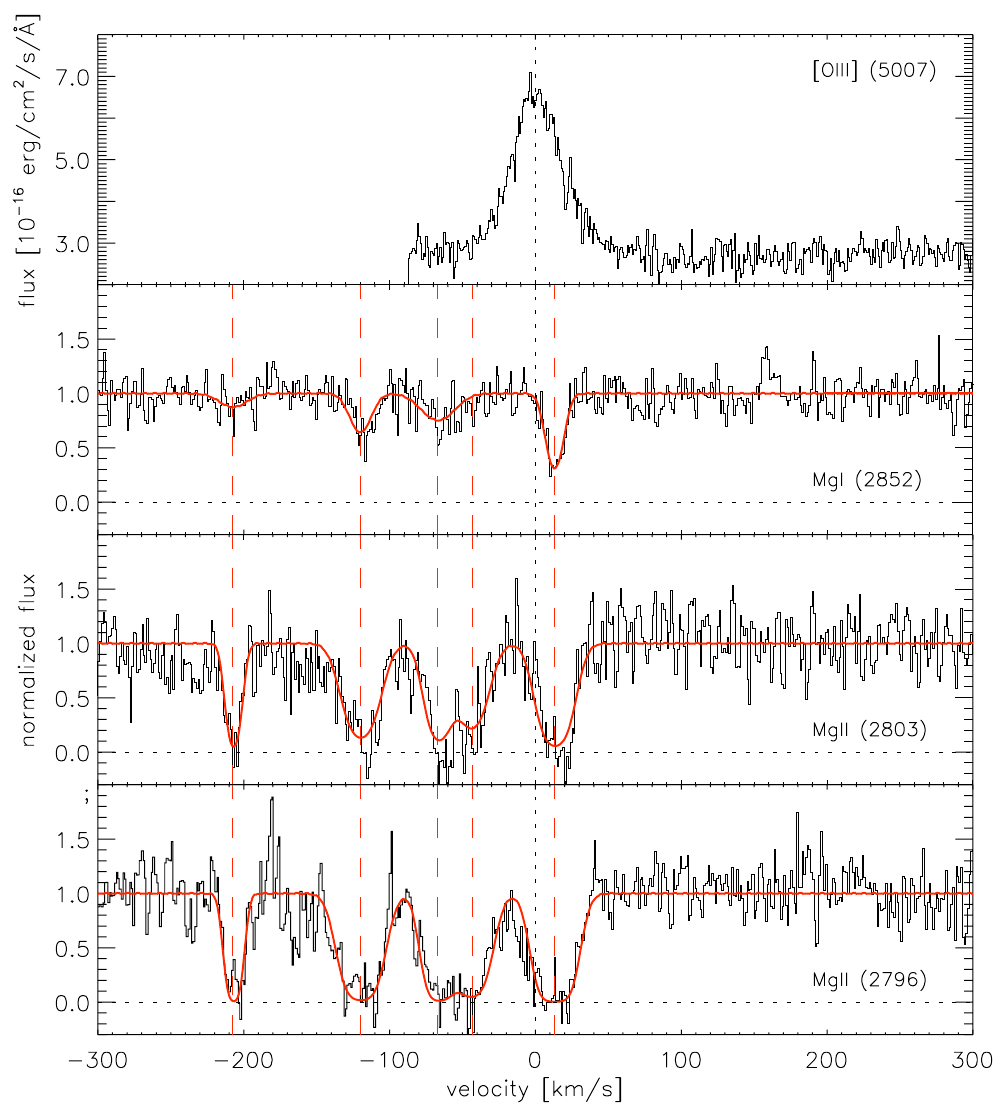}~~~~~~~~~~~~~
\includegraphics[height=8.3cm]{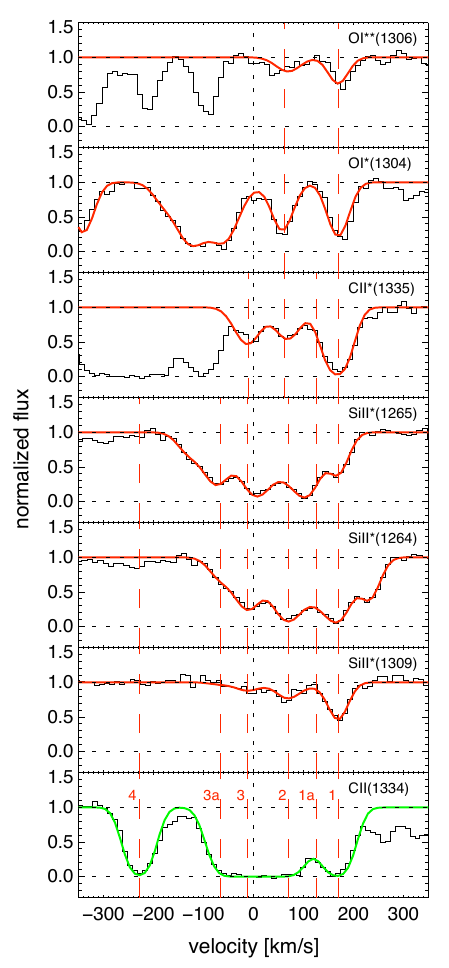}
\caption{Velocity components in the afterglow spectra of GRB 030329 compared to a resolved emission line from the host galaxy (left plot) and those of GRB 060206 showing the fit to the finestructure lines compared to the resonant transition CII in the bottom panel (right plot).}
\end{figure}

\paragraph{GRB 030329}
Its redshift of $z=0.1686$ allowed us to observe both Mg I \& II absorption and the full range of nebular emission lines from the host superimposed on the afterglow spectrum observed on March 29 and April 2 2003 using UVES/VLT \citep{Thoene07}. Five narrow absorption components spanning 250 km/s are blueshifted compared to the resolved host emission lines (FWHM: 50 km/s), that reflect the rotation of the galaxy (Fig. 1). The position of the absorption compared to the emission lines puts them clearly outside the relatively small host. The nonvariability of Mg I between the two epochs leads to a minimum distance of 560 pc to the burst for the redmost component. An outflow caused by a starburst wind is therefore a likely origin for the absorption structures.

\paragraph{GRB 060206}
With z$=$4.084 at the opposite end of the redshift distribution, this burst was observed with ISIS/WHT 1.6\,h after the burst \citep{Fynbo06a, Thoene08b}. Four main velocity systems span a range of 417 km/s and three of them are also detected in the finestructure transitions (Fig. 1), unprecedented in GRB afterglow spectra. The ratio between finestructure and ground-state transitions decreases from the red to the bluer components which is consistent with UV pumping of the transitions by the radiation field of the GRB, but collisional excitation is not excluded. In the case of UV pumping, we can infer a distance of the absorbing material to the burst of about 1 to 8 kpc for the different components. This, together with the large velocity spread, makes an origin from rotation of the galaxy unlikely; a starburst wind is also the preferred scenario.


\section{Spatially resolved hosts - the nearby universe}
The investigation of spatially resolved stellar populations in nearby GRB hosts received increased attention from the debate on two SN-less GRBs \citep[e.g.][]{Fynbo06b} as it was suggested that they could have actually belonged to the short burst population. Studying the immediate environment is another way of getting information about the composition of the GRB progenitor. Both the host of GRB 060505, one of the SN-less GRBs, and the host of GRB 980425 / SN 1998bw were spiral galaxies with the GRB found inside a star-forming region in a spiral arm. A comparison between the stellar populations of the GRB sites in both hosts might provide more information on the SN-less GRB progenitors.

\begin{figure}
\includegraphics[height=4.0cm]{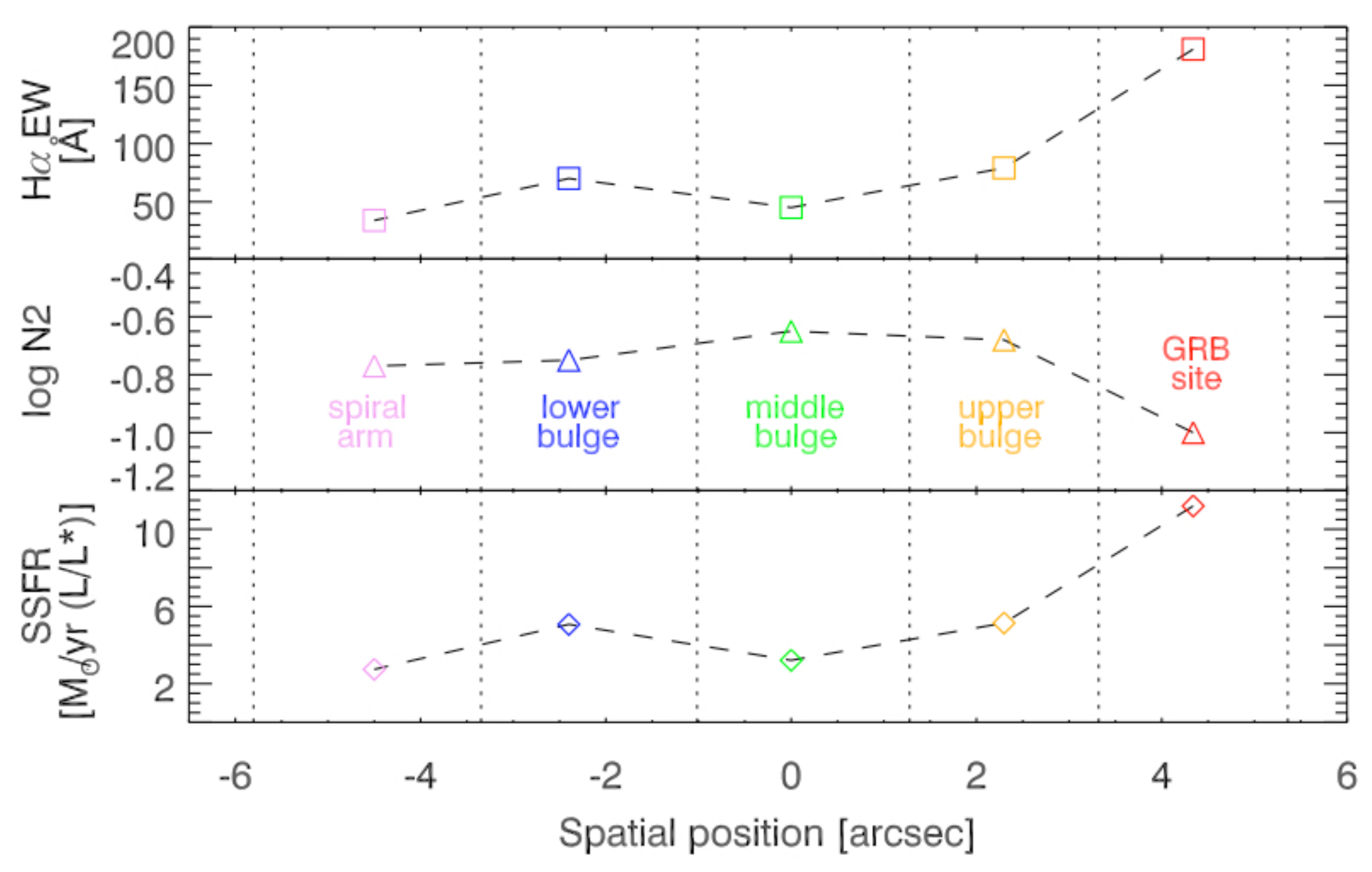}~~~~~~~
\includegraphics[height=4.1cm]{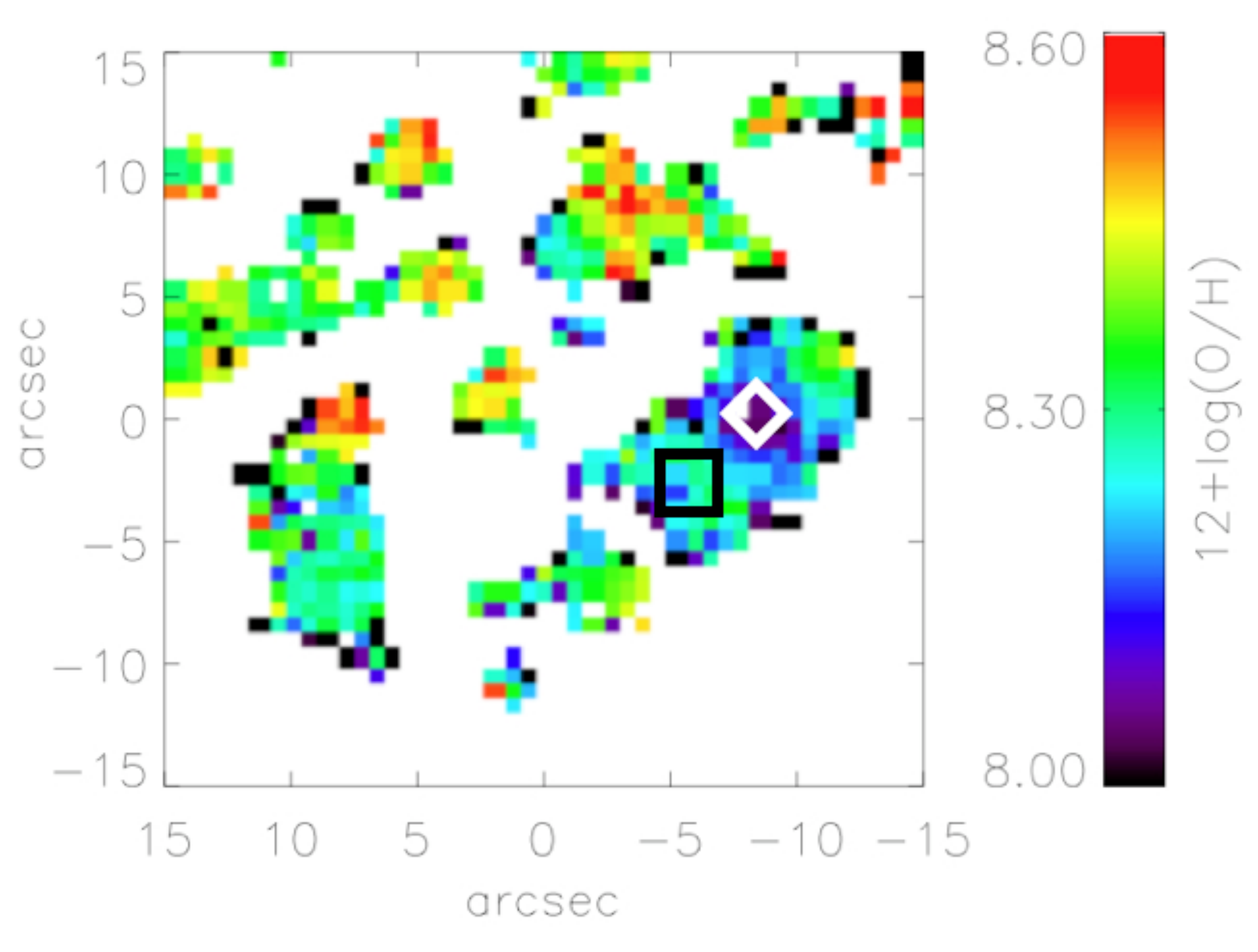}
\caption{Spatially resolved properties of the GRB 060505 host (left panel) and metallicities in HII regions of the GRB 980425 host (right panel), the white diamond and the black square mark the WR and SN/GRB regions respectively.}
\end{figure}

\paragraph{GRB 980425 - a SN GRB}
The host of GRB 980425 was the first one observed with an IFU on VIMOS/VLT providing a spatial resolution of 0.27 kpc (Christensen et al. in prep.). This gives us a detailed map of the metallicity, SFR, extinction and stellar population ages. The GRB/SN region, a smaller star-forming region 0.8 kpc from another SF region with Wolf-Rayet (WR) features detected, turns out to not being very different from other SF regions in this spiral galaxy. The metallicity of the GRB/SN site is not very low though still subsolar (Fig. 2), whereas a low metallicity is suggested for a GRB to occur \citep{WoosleyHeger}. Concerning metallicity and age of the stellar population, the GRB site still lies in the range of other nearby GRB hosts.

\paragraph{GRB 060505 - a SN-less GRB}
We took a longslit spectrum of this host with FORS2/VLT across the galaxy, divided into 5 regions \citep{Thoene08a}. {\it HST} images pinpoint the GRB to a large SR region in a spiral arm and the FORS2 analysis showed it to be a young, highly starforming region with relatively low metallicity whereas the rest of the host and has a much older stellar population and solar metallicity. This makes the burst very likely to be connected to the collapse of a massive star. We have furthermore started an analysis of the kinematics with the first high-resolution data from a GRB host taken with HIRES/Keck (Th\"one et al. 2008c in prep.) deriving e.g. a detailed rotation curve for this galaxy. A detailed mapping of the galaxy is planned with an approved proposal for VIMOS/IFU.

\section{Discussion and conclusions}
We presented two examples of high-resolution afterglow spectra containing metal ground state and finestructure transitions with velocity spreads of a few hundred km/s. Those might probe cold clumps in a starburst wind indicating that starburst winds might have been more common in the early universe. This scenario is supported by 1) the position of the absorption and host emission lines, 2) the nonvariability of neutral species and a distance to the burst derived and 3) the presence and abundance ratio of finestructure lines which can also be transformed into a distance scale from the GRB. The high velocity components found in some bursts are still hard to explain with being connected to winds from the progenitor star, but might infact be caused by foreground galaxies.\\
Nearby GRB hosts allow us to resolve the stellar population around the GRB. The results tell us that \emph{global properties derived for GRB hosts should be considered with caution}. The hosts of GRB 980425, a GRB associated with a SN, and GRB 060505, a SN-less GRBs, had the GRB lying in a SF region in the host with comparable properties to those of other long-duration GRB hosts; however not if one considered the galaxy as a whole, this would not have been the conclusion. Another very recent nearby event, SN 2008D / XRF 080109 in NGC 2770, might provide another missing link for possible deaths of massive stars and a possible connection to the properties of their stellar population.


 %

\bibliographystyle{aipproc}   

\end{document}